\documentclass[aps,pre,reprint,superscriptaddress,showkeys,floatfix]{revtex4-2}
\usepackage[utf8]{inputenc}
\usepackage{amsmath}
\usepackage{amssymb}
\usepackage{amsfonts}
\usepackage{graphicx}
\usepackage{siunitx}
\usepackage{color}
\usepackage{soul}
\usepackage{multirow}
\usepackage[colorlinks=true,citecolor=blue,urlcolor=red]{hyperref}
\usepackage[normalem]{ulem}

\newcommand{\fluc}{\mathrm{fluc}}
\newcommand{\mn}{\mathrm{min}}
\newcommand{\mx}{\mathrm{max}}
\newcommand{\diss}{\mathrm{diss}}
\newcommand{\avg}{\mathrm{avg}}
\newcommand{\app}{\mathrm{app}}

\newcommand{\melt}{\mathrm{melt}}

\newcommand{\NESS}{\mathrm{NESS}}
\newcommand{\EQ}{\mathrm{EQ}}

\renewcommand{\d}{\mathrm{d}}

\begin{document}

\title{Thermal noise of a cryo-cooled silicon cantilever locally heated up to its melting point}

\author{Alex Fontana}\affiliation{Univ Lyon, ENS de Lyon, Univ Claude Bernard Lyon 1, CNRS, Laboratoire de Physique, F-69342 Lyon, France}
\author{Richard Pedurand}\affiliation{Univ Lyon, ENS de Lyon, Univ Claude Bernard Lyon 1, CNRS, Laboratoire de Physique, F-69342 Lyon, France}
\affiliation{Laboratoire des Mat\'eriaux Avanc\'es-IP2I, CNRS, Universit\'e de Lyon, F-69622 Villeurbanne, France} 
\author{Vincent Dolique}\affiliation{Univ Lyon, ENS de Lyon, Univ Claude Bernard Lyon 1, CNRS, Laboratoire de Physique, F-69342 Lyon, France}
\author{Ghaouti Hansali}\affiliation{Ecole Nationale d’Ingénieurs de Saint-Etienne (ENISE), F-42100, Saint-Étienne, France}
\affiliation{Laboratoire des Mat\'eriaux Avanc\'es-IP2I, CNRS, Universit\'e de Lyon, F-69622 Villeurbanne, France} 
\author{Ludovic Bellon}\email{Corresponding author: ludovic.bellon@ens-lyon.fr}\affiliation{Univ Lyon, ENS de Lyon, Univ Claude Bernard Lyon 1, CNRS, Laboratoire de Physique, F-69342 Lyon, France}
\date{\today}

\begin{abstract}
The Fluctuation-Dissipation Theorem (FDT) is a powerful tool to estimate the thermal noise of physical systems in equilibrium. In general however, thermal equilibrium is an approximation, or cannot be assumed at all. A more general formulation of the FDT is then needed to describe the behavior of the fluctuations. In our experiment we study a micro-cantilever brought out-of-equilibrium by a strong heat flux generated by the absorption of the light of a laser. While the base is kept at cryogenic temperatures, the tip is heated up to the melting point, thus creating the highest temperature difference the system can sustain. We independently estimate the temperature profile of the cantilever and its mechanical fluctuations, as well as its dissipation. We then demonstrate how the thermal fluctuations of all the observed degrees of freedom, though increasing with the heat flux, are much lower than what is expected from the average temperature of the system. We interpret these results thanks to a minimal extension of the FDT: this dearth of thermal noise arises from a dissipation shared between clamping losses and distributed damping. 
\end{abstract}

\maketitle

\section{Introduction}
Thermal noise is a phenomenon shared by all systems with a non-zero temperature. It is generated by the energy exchanges between the system and the surrounding environment. In equilibrium, it results in fluctuations of the observables of the system, with an amplitude proportional to the equilibrium temperature. While these fluctuations usually go unnoticed due to their intrinsically small amplitude, they become salient in an increasing number of applications:
in biology they are paramount for bioelectro-magnetism~\cite{Vincze2005} and survival of cells in vitro~\cite{Johnson1972}, in microelectromechanical systems (MEMS) they often limit the sensitivity~\cite{Mohd2009}, and in ground-based Gravitational Waves Detectors (GWD) they prescribe the ultimate resolution~\cite{Harry2006}. Their understanding is thus fundamental.

The Fluctuation-Dissipation Theorem (FDT) stands as the fundamental tool for thermal noise estimations \emph{in equilibrium}. This hypothesis cannot be assumed in many cases: examples range from living systems~\cite{Gupta2017}, aging materials~\cite{Buisson2004} and systems subject to a heat flux~\cite{Monnet2019,Conti2013}. The research of possible non-equilibrium effects on the thermal noise of the test masses employed in GWD recently became a prolific subject~\cite{Conti2013,Komori2018}. Often \emph{higher} fluctuations with respect to equilibrium are expected~\cite{Zhang1998,Conti2013}, in concordance with theoretical predictions such as the Harada-Sasa relation~\cite{Harada2006}. On the other side, we have shown in previous studies that a \emph{lack} of fluctuations is also possible. A silicon microcantilever is brought in a Non-Equilibrium Steady State (NESS) by heating its tip at hundreds of degrees higher than its base thermalised at room temperature. The system, subject to a strong heat flux along its length, is unaffected by this phenomenon, fluctuating as if it was in thermal equilibrium at room temperature~\cite{Geitner2017, Fontana2020}. These results are then interpreted thanks to a minimal extension of the FDT for a system with a non-uniform temperature, demonstrating how the fluctuations are linked to the spatial distribution of the dissipation. 

In this work we push the aforementioned experiments to the physical limits, imposing almost the highest temperature difference the cantilever can sustain, and thus bringing it as far from stationary equilibrium as possible. To do so, we place the sample in a cryostat at \SI{10}{K} and heat its tip close to the melting point with a focused laser, thus prompting a temperature difference of around \SI{1700}{K}. The interest of this experiment is twofold: from a theoretical point of view, the simple extension of the FDT~\cite{Komori2018} is put to a test at its limits. From an experimental point of view, this system can be considered an important test bench for cryogenic high-precision measurements, metrology~\cite{Bon2018} and the GWs community. Indeed, a part of the experimental efforts are heading towards cryogenics (e.g. KAGRA~\cite{Akutsu2019}) in order to reduce the thermal noise of the test masses and suspensions~\cite{Khalaidovski2014}. The deposited heat may generate a NESS which is then paramount to characterise. Furthermore, future detectors might use silicon for the test masses (e.g. Einstein Telescope~\cite{Punturo2010}), thus the same material as the sample of our experiment. 

In the first part, we show how to estimate the temperature of the system in such conditions, through a calibration and a numerical simulation. We then demonstrate how, for a cantilever similar to the one in~\cite{Geitner2017, Fontana2020}, we retrieve a strong dearth of fluctuations, and we interpret it through an estimation of the dissipation in the system. A discussion concludes this work. 

\section{Methods}
\begin{figure}
\begin{center}
\includegraphics[width=0.37\textwidth]{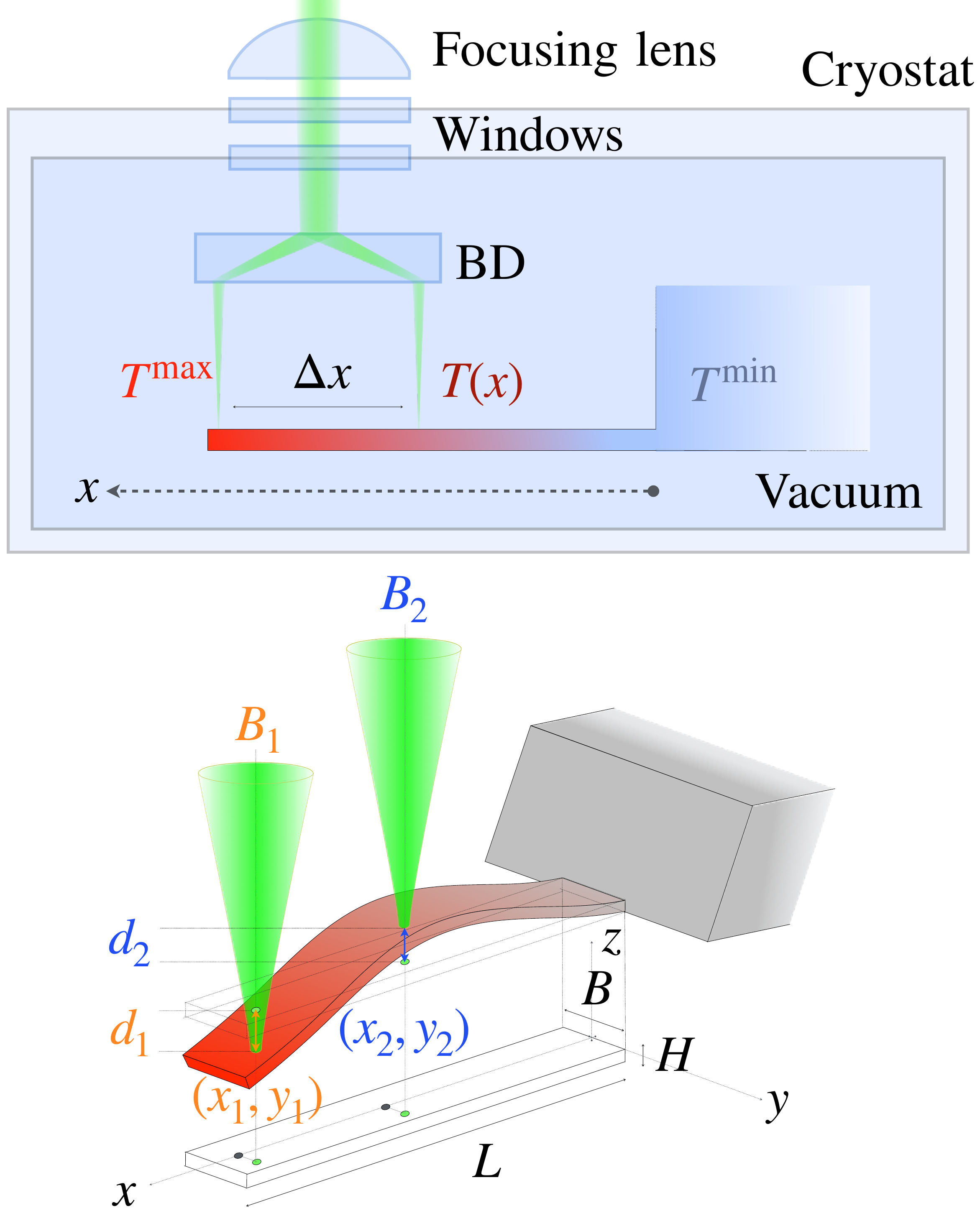}
\caption{(Color online) Experiment setup: the flexion and torsion of a cantilever inside a cryostat are captured thanks to a differential interferometer~\cite{Bellon2002,Paolino2013,PEDURAND2019}. A green laser beam ($\lambda = \SI{532}{nm}$) is divided in two by a beam displacer BD and focused on the cantilever. The interferometer senses the vertical distance $d_1-d_2$ between the beam $B_1$ close the cantilever tip and $B_2$ close to the center. The probing points are separated of $\Delta x = x_1-x_2 = \SI{417}{\micro m}$, $\Delta y =0$. The cantilever, in vacuum at $\SI{e-7}{mbar}$, is monolithically clamped to its macroscopic chip which is thermalised at temperature $T^\mn$. When the laser power is low ($P < \SI{1}{mW}$), we consider the system in thermal equilibrium. When the power is raised ($10$ to $\SI{40}{mW}$), a temperature gradient $T(x)$ along the cantilever arises. }
\label{fig.setupLMA}
\end{center}
\end{figure}

The experimental setup is depicted in Fig.~\ref{fig.setupLMA}. The physical systems consists in a silicon cantilever (OCTOSENSIS micro-cantilevers arrays~\cite{OCTOS}), $L=\SI{1000}{\micro m}$ long, $B=\SI{90}{\micro m}$ wide and $H=\SI{1.1}{\micro m}$ thick. It is monolithically clamped to a macroscopic chip which is kept at $T^\mn\leq\SI{20}{K}$ by a cryostat. The cantilever is placed in a vacuum chamber at $\SI{e-7}{mbar}$. The measuring instrument is the CryoQPDI~\cite{PEDURAND2019}, a Quadrature-Phase Differential Interferometer (QPDI)~\cite{Bellon2002,Paolino2013} combined with a cryostat. A laser beam at $\SI{532}{nm}$ is used to measure the thermal fluctuations of the cantilever. At the same time it acts as a heater when its power $P$ is increased, due to its optical absorption. It is split into two parallel beams by an aberration corrected beam displacer inside the cryostat. The beams are both focused on the cantilever surface with a spot radius $R_0=\SI{3}{\mu m}$. By design, the distance between the two spots is fixed at $x_1-x_2 = \SI{417}{\mu m}$. Since we focus one spot close to the cantilever's free end, the second one is roughly in the center along its length. Across its width, both beams are placed off axis at $y_1=y_2=\SI{37}{\mu m}$.

\begin{figure}
\begin{center}
\includegraphics[width=\columnwidth]{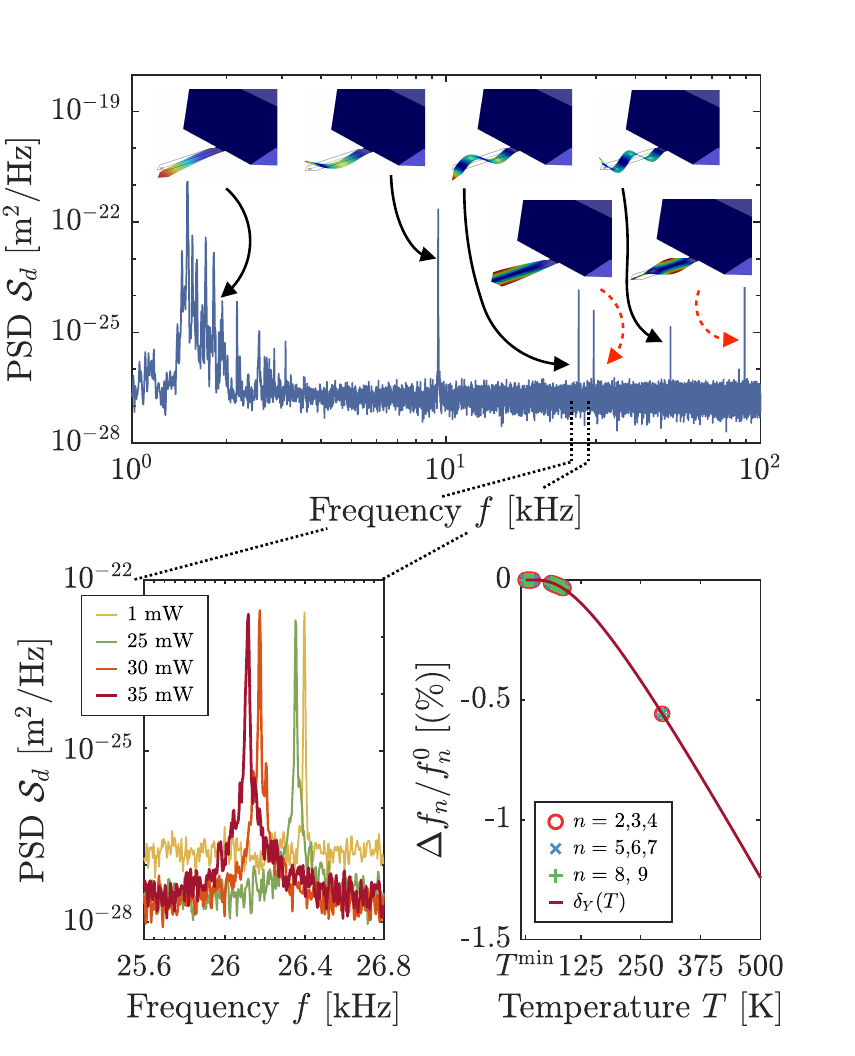}
\caption{(Color online) Experimental thermal noise spectra. (a) The Power Spectrum Density (PSD) of $d$ displays various resonance peaks, which we can distinguish between the flexural ones  and torsional ones, respectively indicated by black solid arrows and red dashed arrows. The first resonance $n=1$ is plagued by low-frequency noise and is thus discarded in the analysis. (b) Increasing the laser power causes a redshift in the resonance frequencies of the modes (in this case $n=3$) as the Young's modulus of silicon decreases with temperature. (c) We measure at equilibrium ($P<\SI{1}{mW}$) the frequency shift with respect to the lowest temperature, when the temperature of the cryostat slowly increases up to room temperature. This calibrates the Young's modulus dependency of temperature $\delta_Y(T)$ through Eq.~\ref{eq.dffgen} with $T$ uniform.}
\label{fig.Spectrum}
\end{center}
\end{figure}

The QPDI senses the optical phase difference between these two beams, which can be swiftly expressed as a vertical difference $d=d_1-d_2$. Up to a geometrical factor dependent of the resonant modes shape, $d$ is sensitive to the flexural deformations (denoted from now on by their amplitude $\delta_n$, with $n$ the mode number) and the torsional ones (denoted by $\theta_m$, with $m$ the mode number). The Power Spectrum Density (PSD) of $d$, plotted in Fig.~\ref{fig.Spectrum}, shows the lowest frequency resonances of the cantilever. Up to 9 flexural and 7 torsional modes are measurable in the experiment. Due to experimental constraints, some are excluded from the analysis: mode $n=1$ is affected by low-frequency external noise (see Fig.~\ref{fig.Spectrum}) and the amplitudes of modes $n,m=5$ are too low due to their vicinity to a node in sensitivity ($d_1$ and $d_2$ are affected likewise by these modes). 

During the measurement, the pulse tube of the cryostat must be turned off: the vibrations it creates during operation are too high for the sensitive thermal noise measurement we perform. We rely on the thermal inertia of the sample holder (lead loaded) to maintain a quasi steady state: the temperature drift is only $\SI{0.18}{K/min}$. Each time $T^\mn$ reaches $\SI{20}{K}$, we turn off the thermal noise measurement and cool down to below $\SI{10}{K}$ before starting a new acquisition. Equilibrium (EQ, with $P<\SI{1}{mW}$) and non-equilibrium (NESS, with $P=10-\SI{40}{mW}$) measurements are alternated to get rid of any drift issue. In addition, we randomise the order of the laser powers and often change the probing point thus shielding the results from particular modifications of the material. Several measurements performed on the same sample also ensure reproducibility~\cite{Fontana2020PhD}.

\subsection{Temperature}

When heating the cantilever with the laser beams, we create a temperature profile $T(x)$ along the cantilever length, which is non-linear: not only there are two heating points, but the thermal conductivity of silicon span 3 orders of magnitude between $\SI{10}{K}$ and the melting point $T^\melt=\SI{1687}{K}$. It is thus not simple to describe the temperature with a single observable. However, for each power $P$ of the laser, $T(x)$ spans from $T(0)=T^\mn$ to $T(L)=T^\mx$, and its average value $T^\avg=\int_0^L T(x) \d x /L$ is used to grasp the out-of-equilibrium character of the system. In this section we describe how to evaluate $T^\avg$ from the measurement.

The resonance frequencies $f_n$ are sensitive to the temperature $T$, mainly through the variation of the Young's modulus $Y$ of silicon. In refs. \onlinecite{Aguilar2015,Pottier2020}, we show that in a first approximation,
\begin{equation}
\label{eq.dffgen}
\frac{\Delta f_n}{f^0_n} = \frac{1}{2} \frac{\int_0^L\d x\, \Delta Y (T(x)) \phi^{\prime \prime}_n(x)^2}{Y^0 \int_0^L\d x\, \phi^{\prime \prime}_n(x)^2},
\end{equation}
where superscript $^0$ stands for the reference value of the quantity at $T^0=\SI{10}{K}$, $\Delta$ stands for the variation of the quantity with respect to the reference one, and $\phi^{\prime \prime}_n(x)$ is the curvature of the normal mode.

When temperature is uniform, the relative frequency shift is independent of the mode number and can be used to calibrate $\delta_Y(T)\equiv\Delta Y (T)/Y^0$.  We perform this calibration of $\delta_Y(T)$ experimentally: the resonance frequencies of the cantilever are tracked with a very low injected laser power so that the system can be considered in equilibrium, while we let the temperature of the cryostat increase slowly from $\SI{10}{K}$ to room temperature \footnote{This calibration was performed on a shorter ($L=\SI{750}{\micro m}$) cantilever of the same manufacturer, but as $\delta_Y(T)$ is independent on geometry, it is valid for the sample used for the thermal noise study.}. Following ref. \onlinecite{Wachtman1961}, we perform a fit of the calibration data with $\delta_Y(T)=c_1T e^{c_2 T}$, leading to $c_1 = (3.68\pm0.04)\times\SI{e-5}{K^{-1}}$ and $c_2 = 196.7\pm\SI{2.7}{K}$. $\delta_Y^{-1}$ can then be used as a thermometer: from a measured frequency shift, one can then deduce the apparent temperature of the cantilever with $T^\app_n \equiv \delta_Y^{-1}(2 \Delta f_n/f_n^0)$. In thermal equilibrium, $T^\app_n$ corresponds to the actual temperature of the cantilever for all $n$. If there is a temperature profile $T(x)$, $T^\app_n$ represents the apparent temperature one would read from such a thermometer. Interestingly, when the mode number is large ($n>5$), curvature is mostly distributed all cantilever long and $T^\app_n$ approximates the average temperature of the system $T^\avg$~\cite{Pottier2020}. Therefore, we can experimentally estimate $T^\avg$ in an out-of-equilibrium situation by $\bar T^\app=\delta_Y^{-1}(2 \langle \Delta f_n/f_n^0\rangle_{n=6-9})$, the average of the apparent temperatures of modes 6 to 9.

To further secure our measurement of $T^\avg$ and have an estimation of the full temperature profile, in appendix~\ref{appendix.temperatureg} we numerically compute $T(x)$ solving the stationary heat equation, taking into account the temperature dependency of the thermal conductivity, thermal radiation, and the two heat sources corresponding to laser absorption. For a given absorbed power, we therefore get $T(x)$, from which we infer $T^\avg$, $T^\mx$, and the relative frequency shift (from Eq.~\ref{eq.dffgen} using the calibrated $\delta_Y(T)$). We end up again with a calibration function, giving $T^\avg_\mathrm{sim}(\langle \Delta f_n/f_n^0\rangle_{n=6-9})$. Both calibrations are very consistent, $\bar T^\app$ overestimating $T^\avg_\mathrm{sim}$  by $\SI{40}{K}$ at most. The temperature profile is however very non-linear with a steep rise close to the end, so that $T^\mx$ on the other hand presents large uncertainties due to the unknown parameters of the problem (mainly the precise knowledge of light absorption).

\subsection{Thermal fluctuations}

All the resonances have a high quality factor ($Q_{n,m} \geq 10^3$, see Fig.~\ref{fig.loss}) and are sufficiently apart from each other to be considered as independent oscillators. Up to a geometrical multiplicative factor, the PSD $\mathcal{S}_d$ around each peak can be seen as the one of the specific mode $\mathcal{S}_{\delta_n}$ or $\mathcal{S}_{\theta_m}$ only. The mean square amplitude of the thermal noise $\langle \delta^2_n\rangle$ or $\langle \theta_n^2 \rangle$ can be evaluated by integrating the PSD in a tiny frequency range around the corresponding peak, subtracting the flat background noise contribution. In equilibrium at temperature $T$, the equipartition principles states that:
\begin{equation} \label{eq.EP}
k_n \langle \delta^2_n\rangle = \kappa_m \langle \theta^2_m \rangle = k_B T,
\end{equation}
with $k_n$, $\kappa_m$ the stiffnesses in flexion and torsion, and $k_B$ the Boltzmann constant. 

When the cantilever is out of equilibrium (under a steady heat flux), we define a fluctuation temperature $T^\fluc$ as:
\begin{equation}
\label{eq.Tfluc}
\begin{split}
T^\fluc_n & \equiv \frac{k_{n} \langle \delta^2_n \rangle}{k_B} = \left(\frac{f_n}{f_n^0}\right)^2\frac{\langle \delta^2_n \rangle_\NESS}{\langle \delta^2_n \rangle_\EQ} T^\mn, \\
T^\fluc_m & \equiv \frac{\kappa_{m} \langle \theta^2_m \rangle}{k_B} = \left(\frac{f_m}{f_m^0}\right)^2\frac{\langle \theta^2_m \rangle_\NESS}{\langle \theta^2_m \rangle_\EQ} T^\mn.
\end{split}
\end{equation}
$T^\fluc$ represents the temperature we would associate to the system through the measurement of its fluctuations, be it in equilibrium or not. Indeed, in this latter regime no thermodynamic temperature of the cantilever can be defined, and $T^{\fluc}_{n,m}$ embody the meaningful value of the fluctuation amplitudes. It is noteworthy that this quantity is in principle mode-dependent, contrarily to the equilibrium case (Eq.~\ref{eq.EP}): indeed, every mode, and thus oscillator, can in principle fluctuate at a different temperature. From an experimental point of view, $T^\fluc$ is calculated as the ratio of the amplitude of the fluctuations in a NESS and in an equilibrium state (EQ, low laser power $P<\SI{1}{mW}$), times the temperature of the thermal bath, corrected by the frequency shift (since $k_n = m_\mathrm{eff} (2 \pi f_n)^2$, with $m_\mathrm{eff}$ the effective mass of the oscillator being independent of temperature, $k_n \propto f_n^2$). As mentioned earlier, each NESS measurement is preceded and followed by an EQ measurement, thus canceling most drift issues when computing $T^{\fluc}$. Moreover, using the ratio of amplitudes avoids any tricky calibration step to measure $\delta_n$ or $\theta_m$.

Using an extended equipartition approach for a NESS~\cite{Geitner2017,Komori2018,Fontana2020}, $T^\fluc$ is expected to be the average of the temperature profile $T(x)$ weighed by the normalised energy dissipation profile $w^\diss(x)$:
\begin{equation}
\label{eq.EPNESS}
T^\fluc_{n,m} = \int_0^L \d x  \, T(x) w^\diss_{n,m}(x)
\end{equation}
In this framework, the fluctuations of the cantilever depend on where the dissipation is preponderant, allowing a wide variety of possible results depending on the shape of $w^\diss_{n,m}(x)$~\cite{Geitner2017,Fontana2020}. We discuss this quantity in the next section.

\subsection{Dissipation}
While it oscillates, the cantilever dissipates energy in the surrounding environment. In high vacuum, hydrodynamical damping is efficiently suppressed and dissipation may arise only from the clamping losses and the internal damping, sometimes referred to as viscoelasticity~\cite{Nowick1972}, arising from local defects or thermoelastic damping for example. Dissipation will thus be a function of the position $x$, frequency $f$, and temperature $T$ which may itself depend on $x$. A generic way to describe it is to introduce the loss angle $\varphi_Y(x,f,T)$ (respectively $\varphi_S(x,f,T)$), which corresponds to the phase of the Young's modulus $Y$(respectively $S$, the shear modulus implied for torsion). Since we are dealing with low dissipation ($\varphi_{Y,S} \ll 1$), the real part of the elastic moduli can be considered independent of $x$ and $f$, and for a given mode $n$ only the value of dissipation at the resonance frequency matters: 
\begin{equation}
\begin{split}
Y(x,f,T) & \approx Y^0(1+\delta_Y(T)+i\varphi_Y(x,f_n,T)) \\
S(x,f,T) & \approx S^0(1+\delta_S(T)+i\varphi_S(x,f_n,T))
\end{split}
\end{equation}

Experimentally, we can only probe the global dissipation by measuring the quality factor $Q_{n,m}=1/\varphi_{n,m}$ of the resonances through a Lorenzian fit of the thermal noise PSD (see Fig.~\ref{fig.loss}). This global dissipation is a function of the mode number $n$ and temperature field $T(x)$~\cite{Geitner2017,Fontana2020}:
\begin{equation}
\begin{split}
\varphi_n\{T(x)\} & = \int_0^L \d x \, \varphi_Y(x,f_n,T(x)) \phi_n^{\prime\prime}(x)^2 \\
\varphi_m\{T(x)\} & = \int_0^L \d x \, \varphi_S(x,f_m,T(x)) \phi_m^{\prime}(x)^2
\end{split}
\end{equation}
Hence, the experimental estimation of $\varphi_{n,m}$ does not allow us to retrieve the spatial dependency of the normalised dissipation $w^\diss_{n,m}(x)$, which writes~\cite{Geitner2017,Fontana2020}:
\begin{equation}
\begin{split}
\label{eq.wdiss}
w^\diss_n(x) & = \frac{1}{\varphi_n\{T(x)\}} \varphi_Y(x,f_n,T(x)) \phi_n^{\prime\prime}(x)^2 \\
w^\diss_m(x) & = \frac{1}{\varphi_m\{T(x)\}} \varphi_S(x,f_m,T(x)) \phi_m^{\prime}(x)^2
\end{split}
\end{equation}
Therefore, we cannot directly calculate the right-hand side of Eq.~\ref{eq.EPNESS} in order to estimate a theoretical value of the fluctuation temperature, dissipation-wise. This is possible just if some hypotheses are satisfied, such as the linearity of the temperature profile, which is not the case of this experiment. We discuss in the next section how $\varphi_{n,m}$ is nevertheless a good indicator for the evolution of the dissipation. 

\section{Results}
\begin{figure}
\begin{minipage}{\columnwidth}
\includegraphics[]{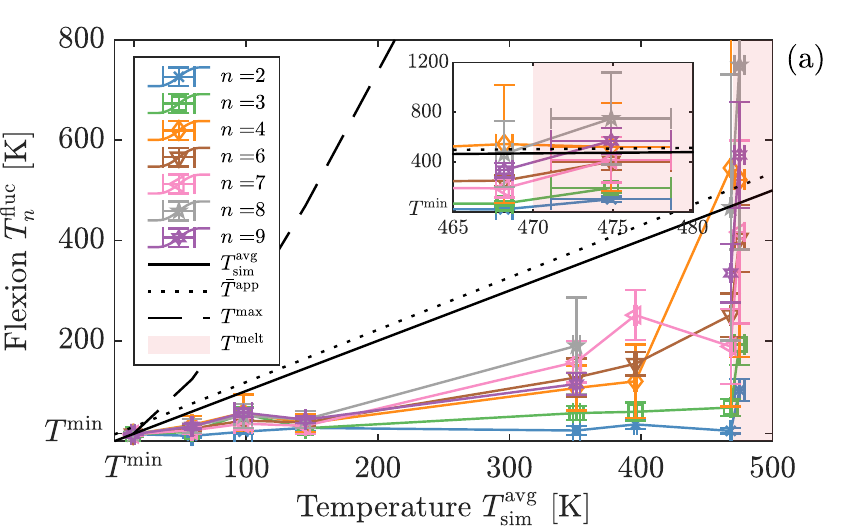}
\end{minipage}
\begin{minipage}{\columnwidth}
\includegraphics[width=\textwidth]{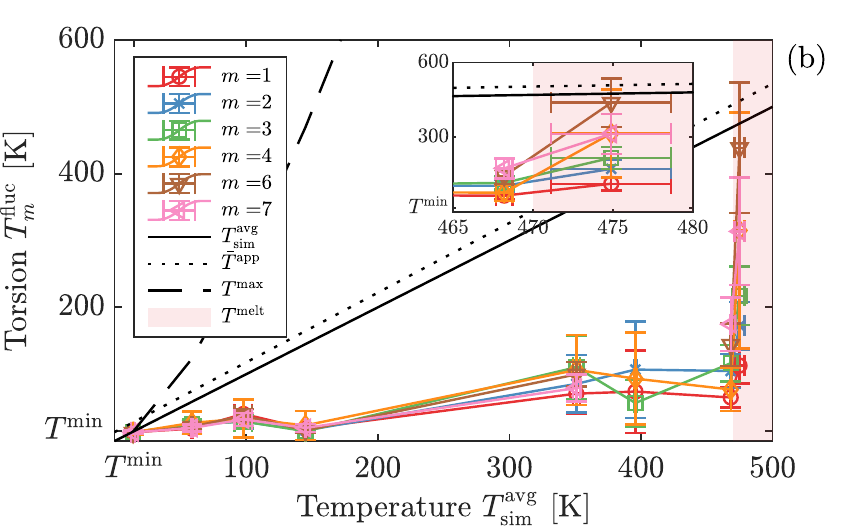}
\end{minipage}
\caption{(Color online) Fluctuation temperature vs. average temperature, for flexion in (a) and torsion in (b). $T^\avg_\mathrm{sim}$ (black solid line) and $T^\mx$ (black dashed line) are evaluated form the measured frequency shift and the calibration function coming from the numerical simulation. The apparent temperature $\bar T^\app$ (black dotted line) is evaluated from the measured frequency shift and the experimental calibration of the Young's modulus temperature dependency. Albeit dependent of $T^\avg$, the amplitude of fluctuations is well below those characteristics temperatures of the cantilever, showing a dearth of thermal noise. The red shaded area on the right covers measurements where at least a partial melting of the cantilever occurred. It is remarkable that in this area $T^\fluc_{n,m}$ greatly increases for most modes of both families. The uncertainties on $T^\avg_\mathrm{sim}$ are discussed in Appendix \ref{appendix.temperatureg}, and the ones on $T^\fluc$ are discussed in the text and in more details in Ref.~\onlinecite{Fontana2020}.}
\label{fig.Tflucs}
\end{figure}

In Fig.~\ref{fig.Tflucs} we show the apparent temperature $\bar T^\app$ and the fluctuation temperatures $T^\fluc_{n,m}$ for all modes as a function of the average temperature of the system $T^\avg_\mathrm{sim}$. At the highest laser power the cantilever begins to melt, which is assessed from camera observations and a reflectivity drop. This indicates that we can reach the highest temperature difference the cantilever can sustain, with $T^\mx -T^\mn \approx \SI{1700}{K}$. For all modes, the fluctuation temperatures are much below the average temperature, except for the highest heating power. The effect is even more striking if we compare $T^\fluc_{n,m}$ with $T^\mx$. This indicates a strong \emph{lack} of fluctuations, as in our earlier experiments on similar cantilevers at room temperature~\cite{Geitner2017,Fontana2020}. It is noteworthy that the $T^\fluc_{n,m}$ show a modest mode dispersion, more pronounced for flexion and almost negligible for torsion. The uncertainties on $T^\fluc$ have two contributions: statistical and systematic. The first is evaluated from the repeated measurement of the thermal noise of the cantilever at the given power. The second takes into account the possibility of the probing point shifting during the measurement, the maximum magnitude of which is estimated to be $dx_1 \equiv dx_2 = \SI{3}{\micro m}$. We discuss this in details in Ref.~\onlinecite{Fontana2020}. Both contributions are equally important in yielding the error bars of fig.~\ref{fig.Tflucs}.

%----------------------------------------------------------------------------------------

In ref. \onlinecite{Fontana2020}, the lack of fluctuations is such that $T^\fluc$ is unchanged when $T^\avg$ increases, leading to the conclusion (through Eq.~\ref{eq.EPNESS}) that the cantilever is dominated by clamping losses. Indeed, if the dissipation is localised at $T^\mn$, it is straightforward to conclude that $T^\fluc \approx T^\mn$. In the present experiment, however, the fluctuations depend on the average temperature. Indeed, we note how they tend to gently increase with $T^\avg$ (except the odd point around $\SI{150}{K}$), reaching up to 10 times the value of $T^\mn$ for the highest heating power. The cantilever cannot therefore be dominated by clamping losses only. The dissipation along the cantilever length should thus have a noticeable contribution.

\begin{figure}
\begin{minipage}{\columnwidth}
\includegraphics[width=\textwidth]{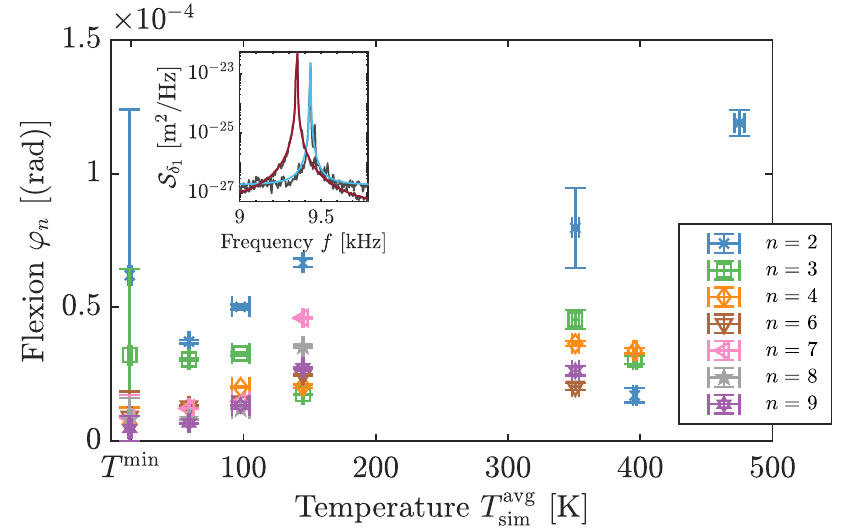}
\end{minipage}
\begin{minipage}{\columnwidth}
\includegraphics[width=\textwidth]{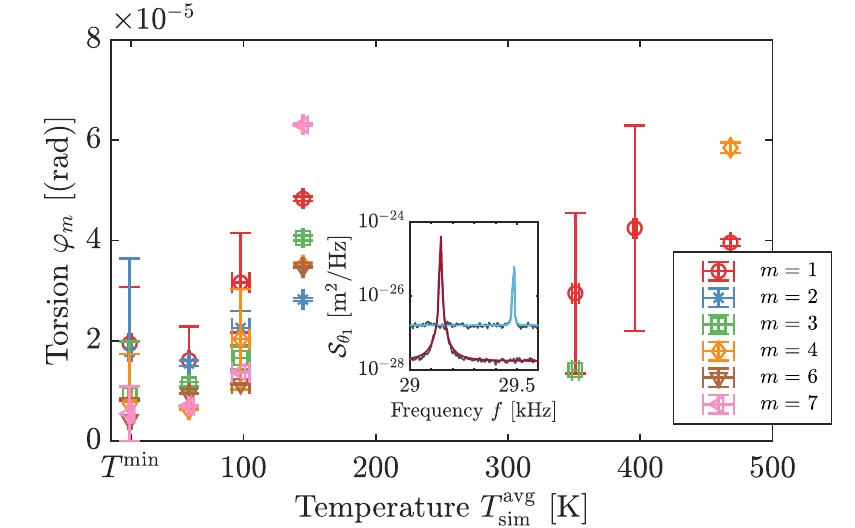}
\end{minipage}
\caption{(Color online) Loss angles $\varphi_{n}$ of flexural modes (a) and $\varphi_{m}$ of torsional modes (b) with respect to the average temperature of the cantilever. Due to experimental contraints and low sample size, the fit of the experimental spectra are often hard to compute, thus we show only the results of fits with a $\chi^2<3$. For most modes, the general trend is an increase of dissipation with increasing temperatures. In order to visually assess this, in the insets we show the peak of the PSD of the second modes in flexion and torsion, widening from low to high temperature, depicted respectively in cyan (light grey) at $T^\avg_\mathrm{sim}=\SI{14}{K}$ and red (dark grey) at $T^\avg_\mathrm{sim}=\SI{474}{K}$ for flexion and $T^\avg_\mathrm{sim}=\SI{145}{K}$ for torsion. The uncertainties on the loss angles are discussed in the text.}

\label{fig.loss}
\end{figure}

A reasonable assumption is that the local dissipation is dependent on the temperature, and will thus introduce a dependence on space in the weighting of $T(x)$ in Eq.~\eqref{eq.EPNESS}. We then expect to measure also a dependence on temperature of the global dissipation of each mode. The measured loss angles $\varphi_{n,m}$ are plotted vs $T^\avg_\mathrm{sim}$ in Fig.~\ref{fig.loss}, confirming this picture. As mentioned, they cannot in general lead to $w^\diss(x)$; nevertheless, they give the qualitative evolution of the dissipation with respect to the average temperature of the cantilever. For all modes, $\varphi_{n,m}$ depend on $T^\avg$, and increase up to 10 times at the highest heating point. It is important to note here that the estimation of $\varphi$ is not trivial in the experiment: due to the slow change of the temperature of the cryostat $T^\mn$ and the fluctuating laser power $P$, the resonance peaks shift during the measurement and artificially enlarge the PSD. Furthermore, the cantilever can sometimes enter in a self-oscillation state, which can inject energy into the resonances altering the results. For these reasons, a careful analysis based on the statistical properties of the PSD is performed~\cite{Fontana2020PhD}, and a large number of spectra are discarded. We choose to show in Fig.~\ref{fig.loss} the results of the fits with a goodness-to-fit $\chi^2<3$ (with 1 being a perfect fit), discarding the others. At each heating power $P$, we fit with a Lorentzian each spectra passing the selection. Each fit provides a measurement of the loss angle $\varphi$ and its uncertainty. Those measurements are then averaged together to compute the final estimation of $\varphi$, and the total uncertainty is calculated as the quadratic sum of the dispersion of the $\varphi$ and of the single uncertainties. In the present experiment, due to the small number of spectra satisfying the applied criteria, the dispersion of the data represents the most important source of error.

The non-trivial profile of $T(x)$ and the unknown $w^\diss(x)$ hinder an estimation of $T^\fluc$ through the extended FDT (Eq.~\ref{eq.EPNESS}). Nevertheless, it is possible to explain the experimental results through some hypothesis on $\varphi(x,f,T)$. We believe this system to be dominated by two main sources of dissipation: clamping and distributed losses. The former is the main source of damping for similar cantilevers at room temperature~\cite{Fontana2020}, and it causes the strong lack of fluctuations we observe. The latter is the responsible of the increase of fluctuations. The loss angle could thus be written as:
\begin{equation}
\label{ }
\varphi_{Y,S}(x,f,T) \approx \varphi^0_{Y,S} (f,T^\mn) \delta^\mathrm{D}(x) + \varphi^1_{Y,S} (x,f,T)
\end{equation}
with $\delta^\mathrm{D}$ Dirac's delta function, $\varphi^0_{Y,S} \approx \SI{e-5}{}$ the loss angle at $T^\mn$ and $\varphi^1_{Y,S}$ an unknown function embedding the evolution of the damping with the temperature and position. With this simple description, we can see that $T^\fluc$ is brought close to $T^\mn$ by the first term, while the second one acts as a correction, becoming important as $T^\avg$ increases. 
%----------------------------------------------------------------------------------------
%----------------------------------------------------------------------------------------
\section{Discussion}
A mechanical system in thermal equilibrium shows no surprises from the fluctuation point of view: all the measurable resonances have an energy content proportional to the temperature of the surrounding thermal bath. When the system is brought to a NESS through a heat flux along its length, on the other side, the thermal noise of the system is not trivial anymore : it depends both of the temperature profile and of \emph{where} the dissipation is localised (Eq.~\ref{eq.EPNESS}). Furthermore, it is in principle possible that different resonances show different results, meaning that the frequency is also a relevant actor. Our system, a silicon micro-cantilever, is thus a suitable test bench in order to characterise the dependency of fluctuations from these parameters, as it is possible to alter their dissipation adding a coating~\cite{Geitner2017}, study a large range of frequencies due to the high number of measurable modes, and greatly vary the temperature. In this experiment, we focus on this last point, exploring the thermal noise of the cantilever between cryogenic temperatures and the melting point of the material. We show how all the measurable resonances show an important dearth of fluctuations, in line with previous experiments on similar samples~\cite{Fontana2020}. Nevertheless, the fluctuations increase with the temperature difference imposed on the system, as does the measured dissipation. Interpreting this thanks to the extended FDT, we conclude that the dissipation profile is non-trivial with clamping losses and distributed damping.

From a theoretical point of view, this work represents an interesting test bench for the minimal extension of the FDT for systems in a NESS~\cite{Komori2018}. Indeed, the cantilever is brought as far from equilibrium as possible, with a ratio larger than 100 between the lowest and highest temperatures, where higher-order corrections of the FDT might in principle be more salient. Not only our results can be perfectly embedded in this framework, but the simultaneous measurement of the damping add support to its validity. Moreover, we see how this is true for the whole frequency range explored ($10-\SI{500}{kHz}$), in which all the modes show a similar behavior. This suggests that the lack of fluctuations is a global property of this cantilever. Experiments such as the one presented in this work represent then a suitable testing bench for out-of-equilibrium thermodynamics: fluctuation theorems and the relative corrections~\cite{Harada2006} and inequalities~\cite{Horowitz2020} can be swiftly put to a test in a simple framework. 

From an experimental point of view, this experiment can be relevant in other fields. For example, the characterisation of the effects of a temperature inhomogeneity can become salient in the noise estimation of micro and nanoresonators~\cite{Cleland2002}. Indeed, the readout laser power needs to be very small in order not to modify the temperature field of the system and thus the amplitude of the fluctuations. Our results show that this condition may be relaxed if the dissipation is localised at the lowest temperature point. Furthermore, as Eq.~\ref{eq.EPNESS} entangles the temperature field with the dissipation field to give the amplitude of the noise, a measurement of two of these quantities yields important information on the third, in cases where its measurement is not possible (e.g. frequency resolution comparable with the width of the resonance). It is similarly possible to perform measurements where we change the probing point along a system and thus test the presence of defects looking at the amplitude of fluctuations: this paves the way to the localisation of single dissipative points. The interest of exploring cryogenic regimes lies in showing how this can be considered possible no matter the temperature. 

As previously mentioned, these results can be useful to the GWs community in characterising the behavior of silicon under a heat flux at low temperatures. The experimental setup of the present work is conceived explicitly to study the temperature dependency of the dissipation of the coatings for the test masses in VIRGO~\cite{PEDURAND2019}. We show here how we can at the same time study possible non-equilibrium effects on the noise, the reduction of which is paramount to increase sensitivity. Doing so with a pure silicon cantilever, we verify how the thermal fluctuations of our sample are weakly dependent on the deposited heat, as they sensibly increase only when the temperature is hundreds of times the one of the cryostat. For this reason, we might expect the deposited heat on the test masses to be less harmful than the equilibrium prediction~\cite{Khalaidovski2014}, fluctuation-wise. It is also important to note that our conclusion for a microscopic system might not hold when we increase in size~\cite{Conti2013}, or when second-order effects in the temperature arise~\cite{Lumbroso2018}. 

To conclude, this work shows how the thermal fluctuations of a micro-cantilever, which base is thermalised at around $\SI{10}{K}$, show a weak dependency on the strong heat flux imposed on the system. This behavior is interpreted thanks to a minimal extension of the FDT, which allows us to link the thermal fluctuations of the cantilever with its dissipation profile. We finally show how the measurement of the global damping is coherent with our theoretical framework. While extended FDT is a valid description for various samples studied in our group~\cite{Geitner2017,Fontana2020}, further studies may comprehend a thorough investigation of exotic dissipation profiles through different geometries, coatings and materials. 
%----------------------------------------------------------------------------------------
\section*{Acknowledgments}
The authors would like to thank L.~Mereni and all the members of the Laboratoire des Matériaux Avnacés for the access to their clean room and to the Cryo-QPDI, and also for their technical support. Financial support from the LABEX Lyon Institute of Origins (Grant No. ANR- 10-LABX-0066) of the Université de Lyon within the program “Investissements d’Avenir” (Grant No. ANR-11-IDEX-0007) of the French government operated by the National Research Agency (ANR) is acknowledged. This work has been also  supported by the Fédération de Physique Ampère in Lyon, and the Mission pour l'Interdisciplinarité of the CNRS.

\section*{Data availability}
The data that support the findings of this study are openly available in Zenodo at \url{https://doi.org/10.5281/zenodo.4696490}~\cite{Fontana-2021-Dataset-PRE}.

\appendix
\section{Temperature simulation}\label{appendix.temperatureg}

In this appendix, we describe the numerical resolution of the heat equation governing the cantilever temperature field and how to use the experimental frequency shifts to evaluate the average and maximum temperature with the help of these simulations.

The heat equation relating the temperature field to the heat fluxes in the problem is strictly speaking a 3D equation. However, since we are interested in length scales larger than $\sim L/10=\SI{100}{\mu m}$ (9 modes in flexion, 6 in torsion), no relevant phenomenon is expected along the thickness $H=\SI{1.1}{\micro m}$. Along the width $B=\SI{90}{\micro m}$, some 2D effects could start being noticeable. In ref.~\cite{Fontana2020PhD}, we show however that if the goal is to estimate $T^\avg$, reducing the problem to 1D yields a difference from the 2D of $5 \%$ at most, which we consider small with respect to other sources of uncertainty.

\begin{figure}
\begin{center}
\includegraphics[width=\columnwidth]{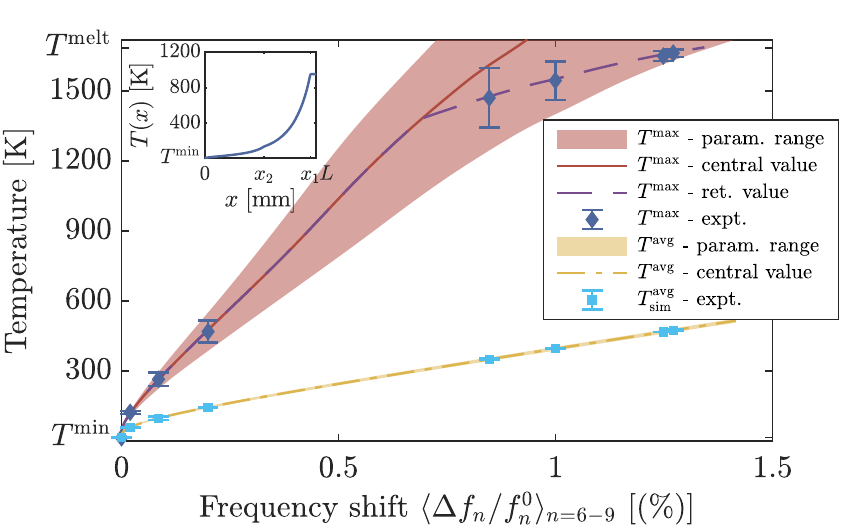}
\caption{(Color online) Estimation of $T^\avg$ and $T^\mx$: from the parametric sweep reported in Table~\ref{table.parameters}, we estimate the possible values of the average and maximum temperature of the cantilever and at the same time the frequency shift for the mode numbers 6 to 9. The relations between these quantities are shown as yellow (light gray) and red (dark grey) curves. The dispersion of $T^\avg$, depicted as a yellow (light grey) shaded area, is small. As a result, the error associated to the experimental value of $T^\avg$ at a given frequency shift, shown as cyan squares is also small. Conversely, the estimation of $T^\mx$ yields a wide parameter range displayed as a red (dark grey) shaded area. No bijective relation is possible, thus we estimate $T^\mx$ as the average of a uniformly distributed variable between the possible values at a given frequency shift. The purple (dark grey) dashed curve represents the retained value of $T^\mx$ for each frequency shift and the blue (dark grey) diamonds are the experimental values. In the inset we show a typical temperature profile $T(x)$, with each of the two laser beams (at $x_1$ and $x_2$) injecting an absorbed power of $\SI{15}{mW}$. We can see how $T(x)$ is highly non-linear and peaked at the heating points.}
\label{fig.Tmax}
\end{center}
\end{figure}

We thus write a stationary 1D heat equation for the cantilever:
\begin{equation}
\label{eq.heatlaw1D}
\begin{split}
\frac{\partial}{\partial x} \Big( \kappa_s\big(T)\big) \frac{\partial T}{\partial x} \Big) + \frac{2 \epsilon_s \sigma_\mathrm{SB}}{H}\left(T^4-{T^\mn}^4\right) + & \\ 
\frac{A_1P_1}{\sqrt{\pi} H B R_0} e^{-2\frac{(x-x_1)^2}{R_0^2}} +  \frac{A_2P_2}{\sqrt{\pi} H B R_0} e^{-2\frac{(x-x_2)^2}{R_0^2}} &=0 
\end{split}
\end{equation}
where $\kappa_s$ is the thermal conductivity of silicon, $\epsilon_s$ its emissivity and $\sigma_\mathrm{SB}$ the Stefan-Boltzmann constant, and $A_iP_i$ the absorbed light power at position $x_i$ ($i=1,2$). The boundary conditions are:
\begin{equation}
\label{eq.BC1D}
\begin{split}
T(0) &= T^\mn\\
\frac{\partial T}{\partial x}(L)&=0
\end{split}
\end{equation}
The first term of Eq.~\ref{eq.heatlaw1D} represents the conduction, the second the radiation and the last ones the two heat sources due to the partial absorption of the laser light. While $\kappa_s(T)$ is tabulated~\cite{Warlimont2018}, the other parameters have large uncertainties the experiment:
\begin{itemize}
\item The nominal thickness H of the cantilever is given by the manufacter with an important uncertainty ($H = 1\pm \SI{0.3}{\micro m} $). Nevertheless, we can deduce its value looking at the flexural resonance frequencies and confronting these values with the Euler-Bernoulli prediction. This gives $H = 1.1 \pm \SI{0.1}{\micro m}$, which is confirmed by scanning electron microscopy images.
\item The emissivity is unknown and it varies greatly at high temperatures~\cite{Jain1971}, where the radiation term is more relevant. A first approximation is to consider $\epsilon_s$ as free parameter (between 0 and 1), independent of the coordinates, to be adjusted. 
\item $T^\mn$ slowly drifts between $\SI{10}{K}$ and $\SI{20}{K}$ during our protocol.
\item Finally, the absorbed power is also unknown, since during the experiment we measure the total injected power $P=P_1+P_2$, with no control over the absorption $A_1$ and $A_2$ (which can be different for each heat source and temperature dependent~\cite{Pottier2021}). It is similarly not possible to know the repartition of the laser power into the two sensing beams, as it could be not equal for $B_1$ and $B_2$. We refer to this balance with $a=A_1P_1/AP$, with $AP = A_1P_1+A_2P_2$ the total absorbed power. We estimate that $a$ can vary for 0.3 to 0.7 in our experiment.
\end{itemize}
Since those parameters are unknown, we then perform a parametric sweep of the aforementioned meaningful quantities, in order to retrieve the family of temperature gradients $\{T(x)\}$ by numerically solving Eq.~\ref{eq.heatlaw1D}. We report the explored range of the parameters in Table~\ref{table.parameters}. For any given set of parameters, we solve the boundary value problem (eqs. \ref{eq.heatlaw1D} and \ref{eq.BC1D}) to extract a numerical solution $T(x)$. One example is shown in the inset of Fig.~\ref{fig.Tmax}, demonstrating the high non-linearity of the profile. As it turns out, $a$ is the most important parameter in prescribing the shape of $T(x)$, and thus $T^\avg$. On the other side, a smaller $H$ or $\epsilon_s$ or a higher total power $AP$ yields a higher $T^\mx$.

\begin{table}[htbp]
\caption{Parameter range for the temperature profile simulations.}
\begin{center}
\begin{tabular}{| c | c | c | c | c | c | c |} 
\hline
& $\epsilon_s$ & $H$ [$\mu$m] & $T^\mn$ [$\SI{}{K}$] & $a$ & $AP$ [$\SI{}{mW}$]  & $n$ \\
\hline
Parameter & \multicolumn{1}{c |}{\multirow{2}{*}{0 - 1}} & \multicolumn{1}{c |}{\multirow{2}{*}{1-1.2}} & \multicolumn{1}{c |}{\multirow{2}{*}{10-20}} & \multicolumn{1}{c |}{\multirow{2}{*}{0.3 - 0.7}} & \multicolumn{1}{c |}{\multirow{2}{*}{1-35}} & \multicolumn{1}{c |}{\multirow{2}{*}{6-9}} \\
range & \multicolumn{1}{c | }{} & \multicolumn{1}{c | }{} & \multicolumn{1}{c | }{} & \multicolumn{1}{c |}{} & \multicolumn{1}{c |}{} & \multicolumn{1}{c |}{} \\
\hline
Central & \multicolumn{1}{c |}{\multirow{2}{*}{0.5}} & \multicolumn{1}{c |}{\multirow{2}{*}{1.1}} & \multicolumn{1}{c |}{\multirow{2}{*}{15}} & \multicolumn{1}{c |}{\multirow{2}{*}{0.5}} & \multicolumn{1}{c |}{\multirow{2}{*}{1-35}} & \multicolumn{1}{c |}{\multirow{2}{*}{7}}\\
value & \multicolumn{1}{c | }{} & \multicolumn{1}{c | }{} & \multicolumn{1}{c | }{} & \multicolumn{1}{c |}{} & \multicolumn{1}{c |}{} & \multicolumn{1}{c |}{} \\
\hline
\end{tabular}
\end{center}
\label{table.parameters}
\end{table}

For each numerical solution $T(x)$, we then compute the average temperature $T^\avg=\int_0^L T(x) \d x /L$, the maximum temperature $T^\mx=\mx(T(x))$, and the relative frequency shift $\langle \Delta f_n/f_n^0 \rangle_{n=6-9}$ through Eq.~\ref{eq.dffgen} (using the experimental calibration for $\Delta Y(T)/Y^0$). All results are finally shown in Fig. \ref{fig.Tmax}.  The solid curve represents the calculated temperature for the central value of the parameters in Table~\ref{table.parameters} and the shaded area all its simulated values.

In order to estimate $T^\avg$ and $T^\mx$ in the experiment, we first average the measured $\Delta f_n/f^0_n$ for $n=6-9$. Then, to each of these values we associate a range of simulated temperatures $\{T^\avg, T^\mx\}$. As we can see, the maximum temperature varies greatly in the simulation. We set an upper bound to its values at the melting temperature $T^\melt$, since we aim to retrieve $T^\mx$ for the measurements where we did not melt the cantilever. In fact, from camera observations and reflectivity estimations we can discern when we damaged the cantilever, hence for the measurements where this is not the case it is reasonable to assume $T^\mx < T^\melt$. The central value of the constrained interval is then the retained value of $T^\mx$, which is depicted as a purple (light grey) dashed curve. The uncertainty associated to $T^\mx$ is then calculated as the standard deviation of the parametric range, taken as if represented by a uniform distribution. Indeed, each value of the parametric range is in principle equiprobable. We perform the same procedure in order to calculate $T^\avg_\mathrm{sim}$ and the respective (small) uncertainty, considering the interval of parameters limited by the upper bound for $T^\mx$. These uncertainties are shown in fig.~\ref{fig.Tmax} for the experimental data as cyan (light grey) squares and blue (dark grey) diamonds. We can see that the numerical simulation gives us a reliable way to estimate the average temperature of the cantilever, and as we see in Fig.~\ref{fig.Tflucs} this is very close to $\bar T^\app$. Conversely, the uncertainty on the unknown parameters hinders the knowledge of $T^\mx$, and the results of the simulation must be taken as an order of magnitude guess.

Finally, the simulations allow us to test the hypothesis that when we shine the cantilever with a low power ($P<\SI{1}{mW}$), the system can be considered close to thermal equilibrium. Since $P$ is measured before the beam is directed towards the vacuum chamber, losses on the optical elements and windows diminish the total intensity that reaches the cantilever. Furthermore, the cantilever absorbs just a part of the shined beam. A conservative guess is to suppose that $AP = \SI{0.5}{mW}$. In this case the simulations give $T^\avg = 15.5 \pm\SI{0.2}{K}$ and $T^\mx = 16.8 \pm\SI{0.5}{K}$ for $T^\mn = \SI{14}{K}$. Therefore, we see how the temperature increase at $P<\SI{1}{mW}$ are very low with respect to the non-equilibrium measurements and the system can safely be considered in thermal equilibrium.

\bibliography{biblio}

\end{document}